\documentclass[11pt]{article}

\usepackage[a4paper,margin=1in]{geometry}
\usepackage[T1]{fontenc}
\usepackage[utf8]{inputenc}
\usepackage{lmodern}
\usepackage{microtype}
\usepackage{amsmath,amssymb,mathtools}
\usepackage{booktabs,tabularx,array,longtable}
\usepackage{graphicx}
\graphicspath{{figures/}}
\usepackage{subcaption}
\usepackage{float}

\newlength{\panelwidth}
\newlength{\panelheight}
\usepackage{xcolor}
\usepackage{enumitem}
\usepackage{csquotes}
\usepackage{tikz}
\usetikzlibrary{arrows.meta,positioning,fit}
\usepackage[colorlinks=true,allcolors=blue!55!black]{hyperref}
\usepackage{orcidlink}
\usepackage[backend=biber,style=numeric-comp,sorting=none,maxbibnames=99]{biblatex}
\newcommand{\Prob}{\mathbb{P}}
\newcommand{\E}{\mathbb{E}}
\newcommand{\Mcal}{\mathcal{M}}
\newcommand{\Dcal}{\mathcal{D}}
\newcommand{\Lcal}{\mathcal{L}}

\title{\textbf{Beyond Zipf's Law: Equifinality and Mechanistic Inference from Scaling Laws}}
\author{Arthur Charpentier\,\orcidlink{0000-0003-3654-6286}
\\[0.45em]
\small Université du Québec à Montréal (UQAM), Canada \& Kyoto University (Japan)\\[0.30em]
\small Corresponding author: charpentier.arthur@uqam.ca}
\date{}

\begin{document}
\maketitle

\begin{abstract}
Scaling laws summarize complex systems through low-dimensional regularities, but the same marginal law can arise from different stochastic dynamics. We examine this ambiguity for Zipf rank--frequency scaling. An i.i.d. finite-Zipf process, a persistent Markov chain, and canonical sample-space reduction (SSR) are constructed to have exactly the same stationary marginal, $p_j=(jH_V)^{-1}$, while a latent-scale mixture produces a similar marginal through aggregation. The first three models therefore hold the population rank distribution fixed while changing temporal organization. Markov dependence shifts the finite-sample distribution of fitted exponents; at moderate persistence, a block-adjusted effective sample size reproduces most of this shift. Excess lag-1 mutual information detects serial dependence relative to a shuffle null, whereas transition direction separates reversible Markov persistence from the directional contraction of SSR. Conditioning on latent scale reveals the aggregation mechanism. Fit-window, alphabet, and sequence-boundary analyses quantify sensitivity to the observation design. These examples separate the population marginal, the finite-sample behavior of a fitted exponent, and temporal structure. Matching a scaling law is therefore a compatibility condition, not a mechanism identifier: discrimination requires observables on which candidate processes make different predictions.
\end{abstract}


\section{Introduction}

Scaling laws compress complex systems into a small number of exponents, but that compression discards information about the dynamics that generated them. The same macroscopic law can arise from different microscopic or mesoscopic processes. Power-law statistics, for example, need not imply criticality \cite{touboul2017}, and temporal correlations can alter sequence-level observables even when the empirical frequency distribution is held fixed \cite{zimmerlin2026}. Zipf's law provides a particularly transparent setting in which to study this many-to-one map from mechanism to observable.

If types are ordered by decreasing frequency, their probabilities are often approximated over some range by
\begin{equation}
    p_{(r)} \propto r^{-\alpha},
    \label{eq:zipf}
\end{equation}
with estimates of $\alpha$ often close to one. We use \emph{canonical Zipf scaling} for $\alpha=1$ and \emph{Zipf-like} for empirical rank--frequency patterns that approximate this form over a stated domain. The relation is documented in language, city sizes, biological data, and other complex systems \cite{zipf1949,newman2005,piantadosi2014}. Its descriptive recurrence has motivated a correspondingly broad set of generative explanations.

These explanations differ substantially. Random segmentation can produce approximate Zipf behavior without semantic organization \cite{miller1957,li1992,conrad2004,perline1996}; reinforcement and innovation generate heavy-tailed frequencies through cumulative advantage \cite{simon1955,gerlach2013,bellina2025}; multiplicative stochastic growth can produce stationary power laws under suitable conditions \cite{gabaix1999,reed2001,yamamoto2014}. Other accounts rely on communicative optimization \cite{ferrer2003,corominasCommunication2011,salge2015}, efficient coding \cite{ferrerCoding2022}, maximum entropy \cite{baek2011,visser2013}, latent heterogeneity \cite{schwab2014,aitchison2016}, or history-dependent restriction of accessible states \cite{corominas2015,thurner2015,mazzolini2018dependency,mazzolini2018ssr}. Several of these mechanisms can coexist, and they need not operate at the same explanatory level.

The inferential question is therefore not whether a candidate mechanism can reproduce a near-one exponent, but which additional observables distinguish mechanisms after that marginal feature has been matched. This distinction is especially relevant for sequence data. Recent work has shown that dependency structure can jointly shape Zipf and Heaps laws \cite{mazzolini2018dependency}, that temporal statistics can distinguish sample-space reduction from other simple generators \cite{mazzolini2018ssr}, and that temporal correlations can decouple type--token growth from the rank--frequency distribution \cite{zimmerlin2026}. Related caution arises in criticality studies, where power-law statistics and scaling functions can occur away from a critical state \cite{touboul2017}.

We use \emph{equifinality} for the broad situation in which distinct processes produce the same observable regularity. The more specific notion of \emph{observational equivalence} depends on what is recorded and how it is summarized: models that agree on one statistic may separate once order, conditioning variables, or longitudinal information are retained. This differs from parameter non-identifiability within a single model and from observation-induced mimicry caused by sampling, segmentation, pooling, or fit-window choice.

Our controlled comparison is built around a stronger condition than approximate agreement of fitted slopes. Three processes---i.i.d. sampling, a persistent Markov chain, and canonical sample-space reduction---share exactly the same stationary finite-Zipf marginal. Their temporal organization can therefore be compared while the one-token population distribution is held fixed. A latent-scale mixture supplies a complementary route in which pooling creates a Zipf-like marginal from non-Zipf conditional distributions. We then examine conditioning, order information, and transition direction using statistics not used for calibration. The aim is model discrimination under specified observation designs, not causal point identification. Learned animal communication is used later as one example of sequence data for which marginal scaling and temporal organization carry different information.

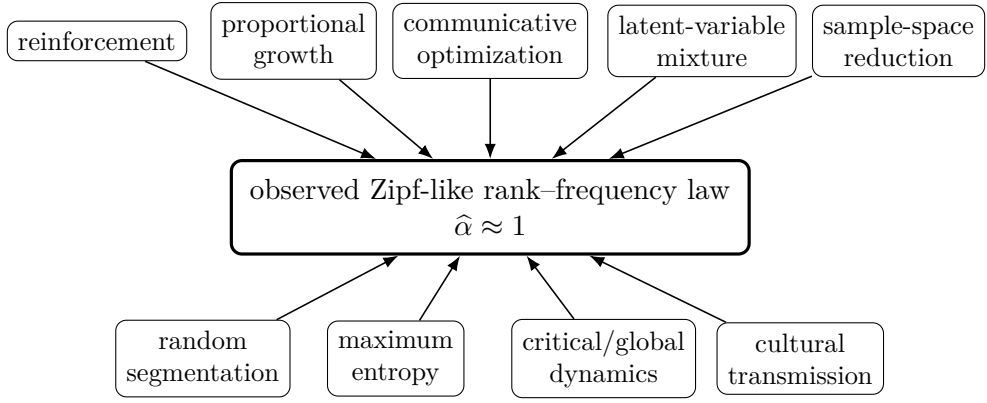
\begin{figure}[t]
\centering
\begin{tikzpicture}[
    mech/.style={draw, rounded corners, align=center, inner sep=4pt, font=\small},
    obs/.style={draw, very thick, rounded corners, align=center, inner sep=7pt},
    arr/.style={-{Latex[length=2mm]}, semithick}
]
\node[mech] (a) at (-5.2,2.0) {reinforcement};
\node[mech] (b) at (-2.6,2.0) {proportional\\growth};
\node[mech] (c) at (0,2.0) {communicative\\optimization};
\node[mech] (d) at (2.8,2.0) {latent-variable\\mixture};
\node[mech] (e) at (5.4,2.0) {sample-space\\reduction};
\node[mech] (f) at (-3.8,-2.2) {random\\segmentation};
\node[mech] (g) at (-1.25,-2.2) {maximum\\entropy};
\node[mech] (h) at (1.5,-2.2) {critical/global\\dynamics};
\node[mech] (i) at (4.1,-2.2) {cultural\\transmission};
\node[obs] (z) at (0,-0.2) {observed Zipf-like rank--frequency law\\$\widehat\alpha\approx 1$};
\foreach \x in {a,b,c,d,e,f,g,h,i} {\draw[arr] (\x) -- (z);}
\end{tikzpicture}
\caption{Zipf's law as an equifinal observable. Distinct mechanisms and generative constructions can produce approximately the same rank--frequency statistic. Distinguishing them requires observations on which the candidate accounts make different predictions.}
\label{fig:equifinality}
\end{figure}

\section{From a scaling pattern to an inferential claim}

``Zipf's law'' can refer to visual linearity on log--log axes, statistical compatibility with a power law over a specified range, or evidence that the exponent is close to one. These claims require different levels of support.

Let $N_i$ denote the count of type $i$ in a sample of size $n$, and let $N_{(1)}\geq N_{(2)}\geq\cdots$ be the counts after empirical re-ranking. We write
\begin{equation}
    \widehat p_{(r)} = \frac{N_{(r)}}{n}.
\end{equation}
A finite-vocabulary generalized Zipf model is
\begin{equation}
    p(r;\alpha,V)=\frac{r^{-\alpha}}{\sum_{j=1}^{V}j^{-\alpha}},
    \label{eq:gzipf}
\end{equation}
where $r$ is the model rank, fixed by the ordering of the theoretical probabilities, and $V$ is the number of admissible types. By contrast, the parenthesized index in $\widehat p_{(r)}$ denotes an empirical rank that can change from one sample to another. The canonical case is $\alpha=1$. Empirical rank--frequency curves often exhibit curvature, finite-size effects, multiple regimes and deviations in the head or tail \cite{piantadosi2014}. Producing a heavy-tailed rank distribution is therefore weaker than explaining why a fitted exponent is near one over a particular domain.

We use three descriptive labels for the relation between a construction and the Zipf target:
\begin{enumerate}[label=(\roman*)]
    \item \textbf{Zipf-generating}: a canonical or structurally central version of the construction yields $\alpha=1$, or converges to it, without choosing the observation window to obtain that value;
    \item \textbf{Zipf-compatible}: the construction generates a broader heavy-tailed or power-law family in which $\alpha\approx1$ occurs for some parameter values but is not structurally imposed;
    \item \textbf{Zipf-mimicking}: a fitted summary approaches one because of finite sampling, ranking, segmentation, aggregation or window choice even though the underlying construction is not canonically Zipfian.
\end{enumerate}
The labels are heuristic and refer jointly to a construction and an observation regime. In finite data, the boundary between compatibility and mimicry is not sharp. We use ``mimicking'' when a near-one fitted exponent is sustained mainly by sampling, preprocessing or a restricted fit window and is not stable under reasonable perturbations of that observation rule. This is a diagnostic description rather than a universal threshold. Random segmentation provides the clearest example in Section~\ref{sec:windows}, while Simon reinforcement supplies a compatible power-law family.

A cross-sectional exponent also carries little direct information about dynamics. Systems can be attracted toward Zipf-like behavior or pass through it transiently \cite{demarzo2021}; models with similar $\widehat\alpha$ can differ in the distributional head and tail, vocabulary growth, sequence structure or temporal evolution. Those differences supply potential discriminating evidence.

Power-law assessment requires an explicit fitted domain, uncertainty quantification and comparison with plausible alternatives; OLS on log--log coordinates is not sufficient \cite{clauset2009}. We retain an OLS slope in Section~\ref{sec:benchmark} only as the familiar one-number summary whose inferential limits the benchmark is designed to expose. A restricted-domain maximum-likelihood estimate on the same ranks provides a second summary, not a stand-alone goodness-of-fit test.

\section{Many roads to Zipf}
\label{sec:taxonomy}

Table~\ref{tab:mechanisms} groups proposed explanations by generative logic and by predictions beyond the marginal rank--frequency curve. The categories overlap: cultural transmission can act through reinforcement, segmentation or compression; coding constraints can implement an optimization principle; latent heterogeneity can coexist with sequential restrictions. The table maps explanatory claims to additional empirical consequences rather than partitioning the literature into mutually exclusive models.

\begingroup
\footnotesize
\setlength{\tabcolsep}{5.5pt}
\setlength{\LTpre}{0.5em}
\setlength{\LTpost}{0.5em}
\begin{longtable}{>{\raggedright\arraybackslash}p{2.4cm} >{\raggedright\arraybackslash}p{4.55cm} >{\raggedright\arraybackslash}p{2.25cm} >{\raggedright\arraybackslash}p{5.05cm}}
\caption{Generative routes to Zipf-like scaling and candidate discriminating evidence. ``Status'' describes the relationship between the mechanism and an exponent near one, not the empirical validity of the mechanism in any particular system.}\label{tab:mechanisms}\\
\toprule
\textbf{Mechanism} & \textbf{Minimal generative idea} & \textbf{Status} & \textbf{Auxiliary prediction / diagnostic} \\
\midrule
\endfirsthead
\multicolumn{4}{l}{\small\itshape Table~\ref{tab:mechanisms} continued}\\
\toprule
\textbf{Mechanism} & \textbf{Minimal generative idea} & \textbf{Status} & \textbf{Auxiliary prediction / diagnostic} \\
\midrule
\endhead
\midrule
\multicolumn{4}{r}{\small\itshape Continued on next page}\\
\endfoot
\bottomrule
\endlastfoot
Random typing / segmentation & Symbols generated independently or with weak dependence are grouped into variable-length units by a boundary process \cite{miller1957,li1992,conrad2004,ferrerElvevag2010}. & Compatible or mimicking & Scaling should be reproducible under null sequences preserving symbol frequencies and the segmentation rule; strong dependence on unit length and boundary statistics. \\
\addlinespace
Reinforcement / cumulative advantage & Existing types are reused with probability increasing with past abundance while innovations introduce new types \cite{simon1955,gerlach2013,bellina2025}. & Power-law family; Zipf in specific regimes & Future increments should depend systematically on current abundance; age, innovation and vocabulary growth provide additional constraints. \\
\addlinespace
Proportional growth & Sizes evolve approximately multiplicatively, with entry, exit or barriers producing a stationary cross-section \cite{gabaix1999,reed2001,yamamoto2014}. & Zipf under identifiable conditions & Conditional growth rates and variances should scale with current size; the model requires a meaningful evolving ``size'' variable. \\
\addlinespace
Communicative optimization & Signal use balances production or speaker cost against ambiguity, listener cost or transmitted information \cite{ferrer2003,corominasCommunication2011,salge2015}. & Compatible; often near optima or transitions & Frequency should covary with measurable cost, ambiguity, surprisal or information, and changes in the trade-off should affect the distribution. \\
\addlinespace
Optimal coding / compression & Probable messages receive less costly codes under coding constraints; rank, probability and code length become linked \cite{ferrerCoding2022}. & Generating under specific coding assumptions & Frequency--length or frequency--cost relations should satisfy the coding solution; predictions extend to other linguistic laws. \\
\addlinespace
Maximum entropy / macroconstraints & Heavy tails arise as least-informative distributions subject to a small set of macroscopic constraints \cite{baek2011,visser2013}. & Power-law family; Zipf for particular constraints & Ensembles preserving only the proposed constraints should reproduce scaling and related aggregate statistics without reproducing unnecessary microstructure. \\
\addlinespace
Latent-variable mixture & Conditional distributions vary across hidden states and their mixture produces an approximately Zipfian marginal \cite{schwab2014,aitchison2016}. & Often robustly Zipf-like & Conditioning on relevant states should weaken the marginal law; deliberate re-mixing should strengthen it. Candidate states include individual, context, sequence length and session. \\
\addlinespace
Critical / global dynamics & Correlated fluctuations near a critical regime, or broader global dynamics, generate scale-free rank structure \cite{schwab2014,touboul2017,morrell2021,ngampruetikorn2025}. & Possible but non-diagnostic & Independent signatures are required: finite-size scaling, susceptibility, correlations or information scaling. Zipf alone cannot identify intrinsic criticality. \\
\addlinespace
Sample-space reduction & Sequence history progressively restricts which states remain accessible \cite{corominas2015,thurner2015,mazzolini2018dependency,mazzolini2018ssr}. & Naturally Zipfian in canonical models & Transition support should contract or be nested with history; randomization that destroys those constraints should weaken the law. \\
\addlinespace
Cultural transmission & Repeated learning and reproduction reshape a signaling system under a transmission bottleneck \cite{arnonKirby2024,arnonWhale2025,kirbyBirdsong2026}. & Empirically compatible; mechanism under study & Statistical structure should change systematically across transmission chains, development or tutor--learner links rather than appearing only in pooled endpoints. \\
\end{longtable}
\endgroup

\subsection{Chance and combinatorics: Zipf-like structure without meaning}

Random typing shows why semantic or communicative organization cannot be inferred from a rank--frequency curve alone. Miller and later Li showed that random character sequences separated into ``words'' can produce an approximate Zipf relation \cite{miller1957,li1992}; Perline connected this construction to randomly indexed sums and the random division of the unit interval \cite{perline1996}, while Conrad and Mitzenmacher generalized it to unequal symbol probabilities \cite{conrad2004}. Such models fail to reproduce richer properties of natural language \cite{ferrerElvevag2010}, but remain useful controls because combinatorics and a boundary process can generate the target statistic without those higher-level properties.

The issue is sharper when the counted units are themselves inferred from sequential statistics. Segmentation then belongs to the observation process, not merely to preprocessing. A proposed mechanism should survive surrogates that preserve low-level symbol frequencies, sequence lengths or transition counts while removing higher-order structure.

\subsection{Reinforcement, innovation and proportional growth}

Simon's model is the canonical cumulative-advantage account: existing types are sampled in proportion to past occurrence while innovations add new types \cite{simon1955}. Modern variants refine the innovation process and connect rank--frequency structure with vocabulary growth \cite{gerlach2013,bellina2025}. What distinguishes reinforcement is its dynamics: current abundance should predict later increments after accounting for exposure and age. A static heavy tail is compatible with reinforcement, but hardly diagnostic of it.

Proportional-growth explanations provide a related but conceptually separate route. In the city-size setting, Gabaix showed how Gibrat-like growth together with stationarity conditions yields Zipf scaling \cite{gabaix1999}; Reed obtained related Pareto and Zipf distributions from stochastic growth observed over random lifetimes \cite{reed2001}. Yamamoto derived a stationary Zipf law from the successive total of a multiplicative stochastic process and related the exponent to symmetry in the random growth rate \cite{yamamoto2014}. These examples show that scaling can emerge from growth dynamics without the sequential restrictions used in the benchmark below.

\subsection{Optimization and coding: explanations with collateral predictions}

Optimization accounts can be tested beyond the exponent because their objective functions imply additional relations. Ferrer-i-Cancho and Sol\'e obtained scaling near a transition between speaker and hearer effort \cite{ferrer2003}; later information-theoretic formulations connected Zipfian structure to communicative efficiency and signal cost \cite{corominasCommunication2011,salge2015}. Efficient-coding approaches derive related frequency and abbreviation laws from constraints on code length and distinguishability \cite{ferrerCoding2022}.

They link frequency to duration or length, production cost, ambiguity, predictability and information. A near-one exponent without those relationships is therefore more problematic for an optimization account than for random segmentation; several linked efficiency relations carry far more evidential weight than the rank--frequency curve alone.

\subsection{Maximum entropy: constraint-based explanation is not process identification}

Maximum-entropy arguments operate at a different explanatory level: given aggregate constraints, they derive the least-committal distribution consistent with them. Baek and colleagues obtain a family of heavy-tailed group-size distributions from minimal information about the number of elements, groups and the largest group \cite{baek2011}; Visser shows directly how a logarithmic constraint yields power-law forms under maximum entropy \cite{visser2013}.

Such a derivation can account for a distribution under stated constraints without specifying how those constraints arose. It can therefore provide a statistical explanation while remaining agnostic about reinforcement, selection or learning. The relevant empirical question is how much of the observed scaling follows from the proposed constraints alone.

\subsection{Latent mixtures and criticality: aggregation as a generator}

Latent-variable models provide a direct route from heterogeneous conditional distributions to a Zipf-like marginal. Schwab, Nemenman and Mehta showed that multivariate systems driven by unobserved fluctuating variables can exhibit Zipf scaling without fine tuning \cite{schwab2014}. Aitchison, Corradi and Latham developed the argument further, demonstrating that a Zipfian marginal can arise naturally after integrating over hidden variables \cite{aitchison2016}. For a latent state $Z$ with distribution $P_Z$,
\begin{equation}
    P(X=x)=\int P(X=x\mid Z=z)\,dP_Z(z).
    \label{eq:mixture}
\end{equation}
Equation~\eqref{eq:mixture} is simply the law of total probability; the substantive mechanism lies in the structure of the conditional family and in how the latent state changes its characteristic scale. In the benchmark below, for example, conditional rank distributions are truncated exponentials whose scale varies broadly across sequences. Pooling over that heterogeneity can display Zipf-like rank structure even when the conditional laws do not.

A direct test is to condition on plausible sources of heterogeneity and ask whether the scaling weakens. In animal communication, $Z$ might include individual identity, tutor, age, social context, recording session, sequence length or within-song position. Conversely, a law that strengthens as increasingly heterogeneous strata are pooled points toward aggregation as part of the explanation.

Criticality provides a closely related caution. Power-law statistics and scaling functions can arise in noncritical stochastic regimes \cite{touboul2017}; latent-variable models can likewise reproduce Zipf-like and other critical-like signatures without intrinsic tuning \cite{schwab2014,morrell2021}. Recent work formalizes a distinction between intrinsic and extrinsic criticality and shows that their observable scaling signatures can overlap \cite{ngampruetikorn2025}. Evidence for criticality must therefore come from diagnostics that are not reducible to the rank--frequency exponent.

\subsection{History-dependent constraints and cultural transmission}

Sample-space reduction offers a particularly transparent sequential construction. If the history of a process progressively restricts the states accessible later, heavy-tailed visitation frequencies can emerge, with canonical constructions yielding Zipf scaling \cite{corominas2015}. Thurner and colleagues applied this mechanism directly to sentence formation, where context progressively narrows the set of words that can follow \cite{thurner2015}. Subsequent work in \emph{Physical Review E} showed how dependency structure can connect Zipf and Heaps laws and how temporal component statistics provide additional discrimination among generators \cite{mazzolini2018dependency,mazzolini2018ssr}. These are precisely the kinds of observables that remain after the marginal frequency law has been matched.

Cultural transmission operates at a higher level. In an iterated-learning experiment, Arnon and Kirby showed that initially unstructured non-linguistic sequences become segmented into recurrent parts and develop increasingly skewed frequency structure as they pass through learners \cite{arnonKirby2024}. Humpback whale song provides a natural example of a culturally transmitted animal signal in which similar statistical structure has been detected \cite{arnonWhale2025}. The Bengalese finch result extends the comparison to a socially learned birdsong system \cite{kirbyBirdsong2026}.

These accounts can operate at different levels of the same process. Cultural transmission may change transition probabilities; learning may favor shorter or more predictable units; reinforcement may operate within transmission chains. Empirical analysis must therefore distinguish the links through which transmission and learning alter lower-level sequence dynamics rather than treat ``culture'' and ``statistics'' as competing labels.

\section{Equifinality, marginal equivalence and model discrimination}
\label{sec:identification}

Equifinality is the broad claim that distinct processes can produce the same observable regularity. We use \emph{marginal equivalence} for the more specific case in which two processes have the same one-token stationary distribution. \emph{Observational equivalence} is defined relative to an observation design and a statistic: two models are equivalent in this sense only if they induce the same sampling law for the recorded summary. Parameter non-identifiability within a single model is a separate issue.

Let $M\in\Mcal$ denote a candidate generator and $\Dcal_n$ the observation design, including sample size, sequence boundaries, segmentation, pooling and fit-window rules. The observed sample follows
\begin{equation}
    X_n\sim P_{M,\Dcal_n},
\end{equation}
and a statistic $T$ has sampling law
\begin{equation}
    \Lcal_{M,\Dcal_n}(T)
    :=\mathcal{L}\!\left(T(X_n)\mid M,\Dcal_n\right).
    \label{eq:samplinglaw}
\end{equation}
If $Q_M$ denotes the stationary one-token marginal, marginal equivalence means
\begin{equation}
    Q_{M_i}=Q_{M_k}.
\end{equation}
It does not imply
\begin{equation}
    \Lcal_{M_i,\Dcal_n}(T)=\Lcal_{M_k,\Dcal_n}(T),
    \label{eq:marginal-not-sampling}
\end{equation}
because dependence and other features of the joint process can affect the finite-sample distribution of $T$. Section~\ref{sec:benchmark} exploits this distinction: the i.i.d., Markov and SSR constructions have the same stationary marginal but different dependence structures.

For a scalar statistic $T_j$, let $\mathrm{AUC}_j(M_i,M_k)$ be the area under the ROC curve obtained when $T_j$ alone orders held-out realizations from $M_i$ and $M_k$. We report
\begin{equation}
    A_j(M_i,M_k)=2\,\mathrm{AUC}_j(M_i,M_k)-1,
    \qquad -1\leq A_j\leq1,
    \label{eq:auceffect}
\end{equation}
and use $S_j=|A_j|$ when only the magnitude of separation matters. Zero corresponds to chance-level ordinal discrimination and unit magnitude to complete separation in the simulated samples. $A_j$ is a univariate rank effect, not a distance between model distributions and not an equivalence test. We report bootstrap intervals without assigning a universal threshold to its magnitude.

Observation design matters because pooling can hide heterogeneity, aggregation can remove order, and a fit window can change the fitted exponent. Additional statistics are informative when plausible alternatives predict different values and the statistics were not used to tune those alternatives. In the primary comparison, the i.i.d., Markov and SSR constructions are fixed analytically. The latent mixture is calibrated on independent simulations using only the OLS exponent: candidate scale ranges are ranked by the absolute deviation of the mean calibration exponent from one, with RMSE and Monte Carlo standard deviation used as tie-breakers. The final latent candidate is evaluated on a prespecified 50-seed calibration panel. Held-out evaluation uses disjoint seeds.

Table~\ref{tab:diagnostics} lists designs that apply directly to the sequence and communication mechanisms considered later. It is not a reduced version of Table~\ref{tab:mechanisms}: proportional-growth, maximum-entropy and criticality accounts require different observables, such as growth increments, conserved constraints or finite-size and susceptibility measures, already noted in the taxonomy. A dash in Table~\ref{tab:diagnostics} means that no comparatively direct prediction is asserted at the level shown; variants of the model class may still imply such a relation.

\begin{table}[t]
\centering
\footnotesize
\caption{Examples of observations that can discriminate accounts sharing a Zipf-like marginal. A check mark denotes a comparatively direct prediction, a circle a potentially informative but non-specific relation, and a dash no direct prediction at the level represented here.}
\label{tab:diagnostics}
\setlength{\tabcolsep}{4pt}
\begin{tabular}{lcccccc}
\toprule
 & Cond. & Shuffle & Transition & Longitudinal & Cost/info & Transmission \\
\midrule
Random segmentation & $\circ$ & $\checkmark$ & $\circ$ & -- & -- & -- \\
Reinforcement & -- & $\circ$ & -- & $\checkmark$ & -- & $\circ$ \\
Optimization/coding & -- & -- & $\circ$ & -- & $\checkmark$ & $\circ$ \\
Latent mixture & $\checkmark$ & $\circ$ & -- & $\circ$ & -- & -- \\
Sample-space reduction & $\circ$ & $\checkmark$ & $\checkmark$ & -- & -- & -- \\
Cultural transmission & $\circ$ & $\circ$ & $\circ$ & $\checkmark$ & $\circ$ & $\checkmark$ \\
\bottomrule
\end{tabular}
\end{table}

Matching a marginal statistic establishes compatibility. Discriminating among generators requires information that was not consumed in producing that match.

\section{A controlled benchmark: identical marginals, different dynamics}
\label{sec:benchmark}

We compare four constructions chosen to separate distinct sources of ambiguity. Three have the same stationary finite-Zipf marginal but different temporal dependence. The fourth produces a similar marginal by pooling non-Zipf conditional distributions. Random segmentation and Simon reinforcement are reserved for the fit-window analysis in Section~\ref{sec:windows}.

\subsection{Design and generative constructions}
\label{sec:benchmark-design}

Each held-out realization contains $n=10^5$ tokens, and all finite-state constructions use $V=5000$. Define
\begin{equation}
    p_j=\frac{j^{-1}}{H_V},
    \qquad
    H_V=\sum_{k=1}^{V}\frac1k.
    \label{eq:finitezipf}
\end{equation}
The four primary constructions are:

\begin{enumerate}[label=(\roman*)]
    \item \textbf{Finite Zipf reference.} Tokens are sampled independently from $p=(p_1,\ldots,p_V)$. For sequence statistics, the realization is divided into blocks of length $L=250$.

    \item \textbf{Zipf Markov control.} Each block starts from $p$ and follows
    \begin{equation}
        P_{ij}=\rho\,\mathbf 1\{i=j\}+(1-\rho)p_j,
        \qquad \rho=0.35.
        \label{eq:markov}
    \end{equation}
    Here $\rho$ is an illustrative persistence parameter, not an estimate from biological data. The chain is stationary under $p$ because $\sum_i p_iP_{ij}=p_j$, and reversible because $p_iP_{ij}=p_jP_{ji}$.

    \item \textbf{Canonical sample-space reduction (SSR).} From state $i>1$, the next state is sampled uniformly from $\{1,\ldots,i-1\}$; from state 1 the process restarts uniformly on $\{1,\ldots,V\}$. The initial state is drawn from stationarity. The stationary distribution is again $p$. For $\pi_j=1/(jH_V)$,
    \begin{equation}
        (\pi P)_j
        =\frac{\pi_1}{V}+\sum_{i=j+1}^{V}\frac{\pi_i}{i-1}
        =\frac1{H_V}\left[\frac1V+\sum_{i=j+1}^{V}\frac1{i(i-1)}\right]
        =\frac1{jH_V}.
        \label{eq:ssrstationary}
    \end{equation}
    where the sum is understood to be zero when $j=V$. SSR is therefore not calibrated to the exponent. Natural cycles between restarts define its primary sequence boundaries.

    \item \textbf{Latent-scale mixture.} Each realization contains 400 sequences of length 250. A scale $S_s$ is sampled once per sequence with $\log S_s\sim\mathrm{Unif}(\log 0.5,\log 8500)$; the implementation uses natural logarithms. Conditional on $S_s=s$, ranks are independent with probability mass
    \begin{equation}
        \Prob(R=r\mid S_s=s)
        =\frac{(1-e^{-1/s})e^{-(r-1)/s}}{1-e^{-V/s}},
        \qquad r=1,\ldots,V.
        \label{eq:latentpmf}
    \end{equation}
    This is the discretized truncated exponential generated by applying $\lceil\cdot\rceil$ to a continuous exponential variate truncated at $V$. The conditional laws are not Zipf laws. The scale range was selected on the independent calibration panel described in Section~\ref{sec:identification}.
\end{enumerate}

Table~\ref{tab:parameters} collects the parameters used in the primary comparison.

\begin{table}[t]
\centering
\small
\caption{Primary simulation parameters. The first three constructions are fixed analytically; only the latent mixture is calibrated to the exponent target.}
\label{tab:parameters}
\begin{tabular}{lll}
\toprule
Construction & Status & Parameters \\
\midrule
Finite Zipf reference & fixed & $V=5000$, $\alpha=1$, block $L=250$ \\
Zipf Markov control & fixed & $V=5000$, $\alpha=1$, $\rho=0.35$, $L=250$ \\
Sample-space reduction & fixed & $V=5000$, stationary start, uniform restart \\
Latent mixture & calibrated & 400 sequences, length 250, $V=5000$, scale $[0.5,8500]$ \\
\bottomrule
\end{tabular}
\end{table}

The primary rank statistic is the OLS log--log slope $\widehat\alpha_{\mathrm{OLS}}$ obtained by regressing $\log\widehat p_{(r)}$ on $\log r$ over
\begin{equation}
    10\leq r\leq500,
    \qquad N_{(r)}\geq5.
    \label{eq:fitwindow}
\end{equation}
We use OLS here as a familiar low-dimensional summary, not as a recommended power-law estimator. A discrete power law is also fitted on the same retained rank set $\mathcal R$. With $N_{\mathcal R}=\sum_{r\in\mathcal R}N_{(r)}$,
\begin{equation}
\widehat\alpha_{\mathrm{RD}}
=\arg\max_{\alpha>0}
\left[
-\alpha\sum_{r\in\mathcal R}N_{(r)}\log r
-N_{\mathcal R}\log\!\left(\sum_{j\in\mathcal R}j^{-\alpha}\right)
\right].
\label{eq:mle}
\end{equation}
We refer to this estimator as the \emph{restricted-domain maximum-likelihood estimate} (RD-MLE) and denote it by $\widehat\alpha_{\mathrm{RD}}$. It conditions on the retained rank domain; it is not a model-selection test against alternative heavy-tailed families.

All primary results use 100 held-out replications. Pairwise AUC intervals are based on 2000 nonparametric bootstrap resamples, resampling the held-out realizations separately within each model.

\subsection{Identical marginals can yield different finite-sample exponent estimates}
\label{sec:markov-finite}

\begin{figure}[t]
\centering
\begin{subfigure}[t]{0.48\textwidth}
    \centering
    \includegraphics[width=\linewidth]{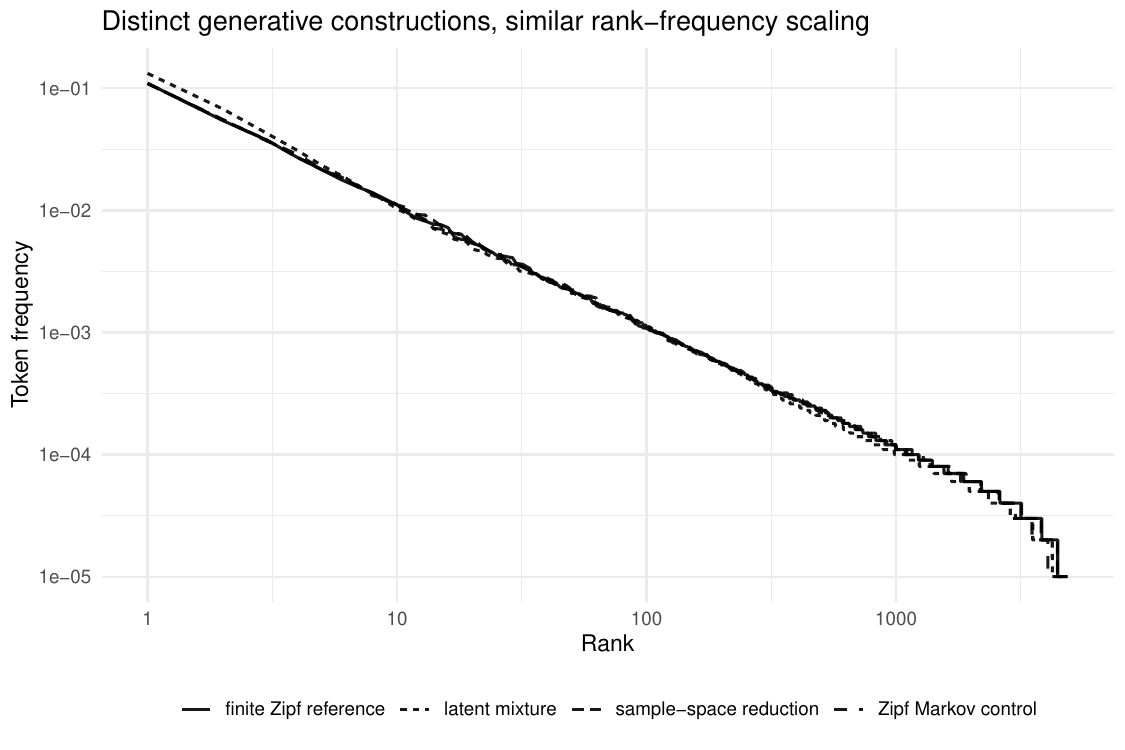}
    \caption{Representative rank--frequency curves.}
\end{subfigure}\hfill
\begin{subfigure}[t]{0.48\textwidth}
    \centering
    \includegraphics[width=\linewidth]{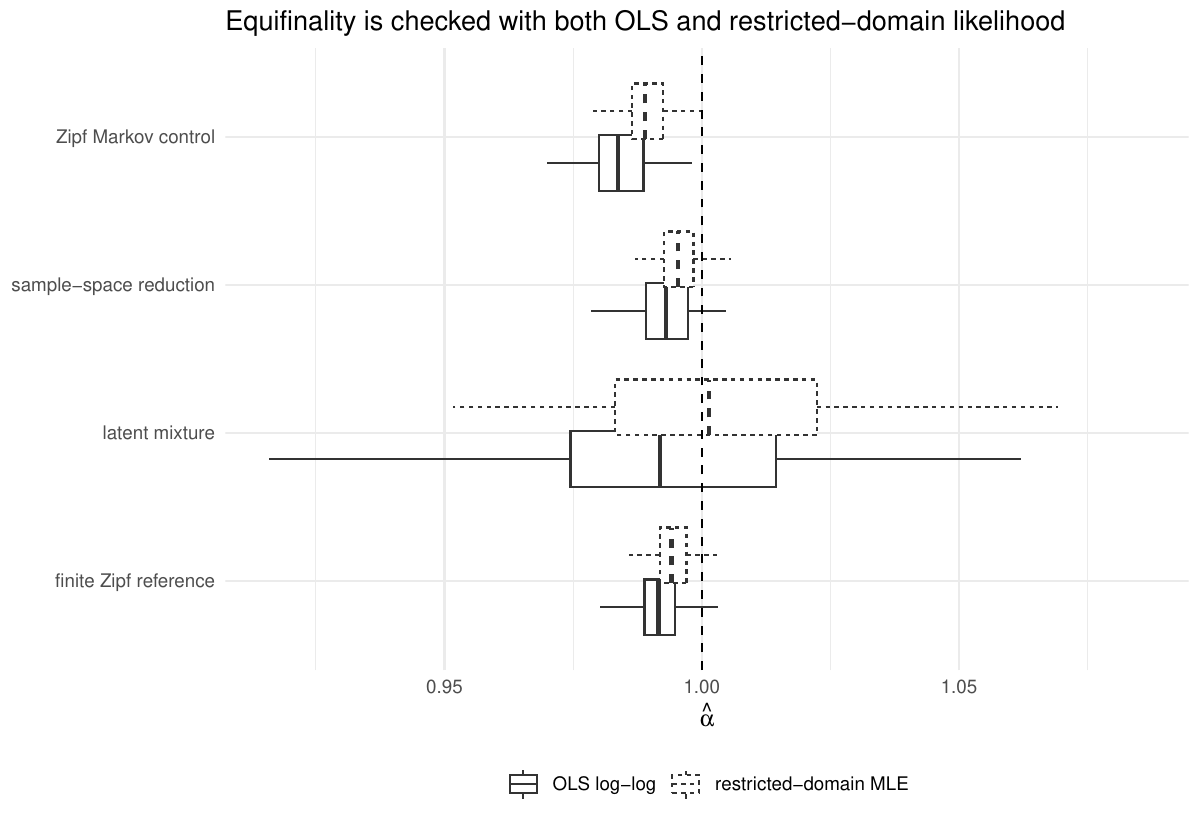}
    \caption{OLS and RD-MLE estimates.}
\end{subfigure}
\caption{Marginal agreement under two exponent summaries. The i.i.d. finite-Zipf reference and canonical SSR are nearly indistinguishable under the primary finite-sample design, while the latent mixture has greater between-replication variability. The Markov process shares the exact stationary marginal of the i.i.d. and SSR constructions but has a shifted finite-sample estimator distribution.}
\label{fig:marginal}
\end{figure}

The rank--frequency curves overlap closely over the primary fitted domain (Figure~\ref{fig:marginal}a). For the i.i.d. reference, Markov control and SSR, this similarity follows from an exact result: all three have the stationary marginal in equation~\eqref{eq:finitezipf}. Figure~\ref{fig:marginal}b shows that their fitted exponent distributions need not coincide.

The i.i.d. reference and SSR have mean OLS exponents 0.992 and 0.993, respectively, and both have mean RD-MLEs of 0.995 (Table~\ref{tab:simresults}). The latent mixture gives 0.994 by OLS and 1.004 by RD-MLE, with much greater between-replication dispersion. For OLS, the signed AUC effects are $-0.012$ (95\% bootstrap interval $[-0.196,0.174]$) for finite Zipf versus latent mixture, $-0.131$ ($[-0.291,0.035]$) for finite Zipf versus SSR, and $-0.016$ ($[-0.200,0.169]$) for latent mixture versus SSR. The corresponding restricted-domain intervals also include zero. Under this design, the exponent has little ordinal discriminating power among these three constructions.

The Markov control has mean exponents 0.984 by OLS and 0.989 by RD-MLE. The signed effects for finite Zipf versus Markov are 0.646 ($[0.530,0.762]$) and 0.588 ($[0.465,0.706]$), respectively. Since the two processes have the same population marginal, this separation concerns the sampling distribution of the estimator under dependence rather than a difference in the population Zipf exponent.

\begin{table}[t]
\centering
\footnotesize
\setlength{\tabcolsep}{4pt}
\caption{Held-out statistics across 100 independent realizations. Entries are means with standard deviations in parentheses. The finite Zipf reference, Markov control and canonical SSR share the same theoretical finite-Zipf stationary marginal.}
\label{tab:simresults}
\begin{tabular}{lccccc}
\toprule
Construction & OLS $\widehat\alpha$ & RD-MLE $\widehat\alpha$ & Final $V_n$ & $I_{\rm excess}$ & $D$ \\
\midrule
Finite Zipf reference & 0.992 (0.005) & 0.995 (0.004) & 4857.8 (12.1) & 0.000 (0.000) & 0.500 (0.001) \\
Zipf Markov control & 0.984 (0.007) & 0.989 (0.005) & 4601.0 (16.2) & 0.188 (0.002) & 0.500 (0.001) \\
Sample-space reduction & 0.993 (0.005) & 0.995 (0.004) & 4855.6 (12.6) & 0.350 (0.001) & 0.946 (0.001) \\
Latent mixture & 0.994 (0.031) & 1.004 (0.029) & 4755.4 (59.9) & 0.000 (0.001) & 0.500 (0.001) \\
\bottomrule
\end{tabular}
\end{table}

\begin{figure}[t]
\centering
\includegraphics[width=0.74\textwidth]{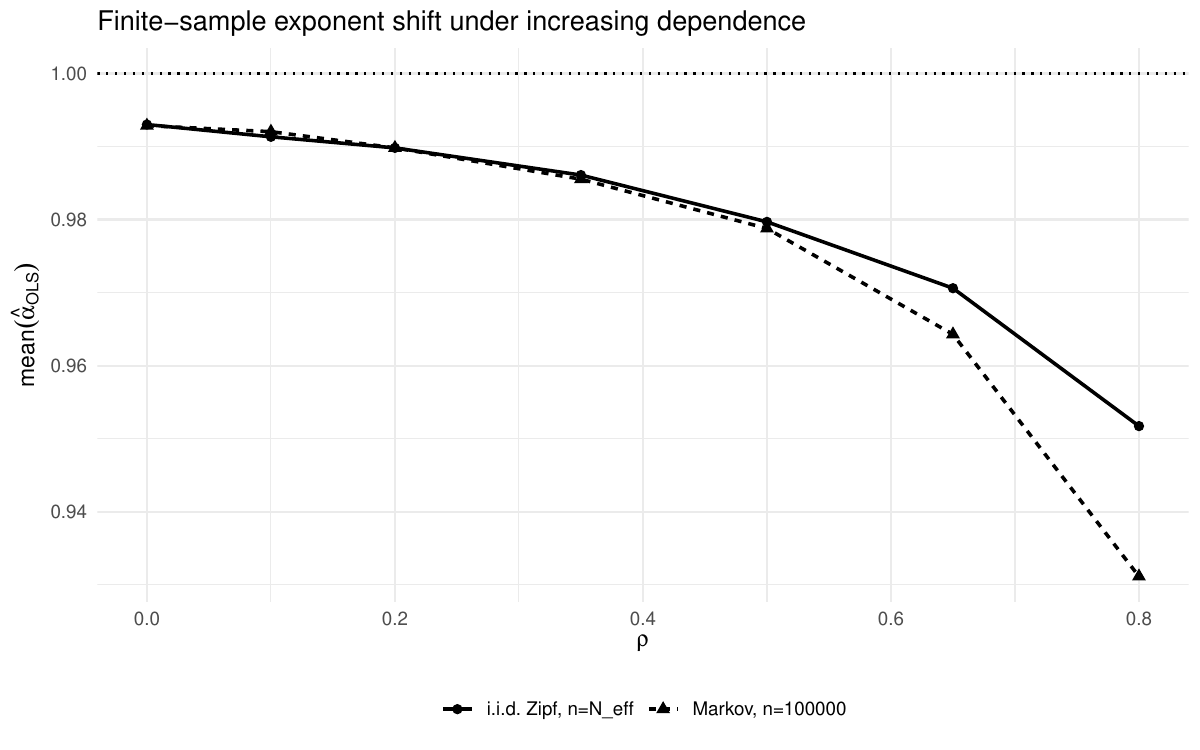}
\caption{Finite-sample exponent shift under increasing Markov persistence. The Markov control uses $n=10^5$; the i.i.d. comparison uses $n=n_{\rm eff}(\rho)$ from equation~\eqref{eq:neff}. The effective-sample-size calculation closely tracks the fitted-exponent shift at moderate persistence and diverges at high dependence.}
\label{fig:neff}
\end{figure}

The Markov shift can be related to the amount of independent information in the sample. Blocks of length $L=250$ are restarted independently from stationarity. For any centered function $f$,
\begin{equation}
    \operatorname{Cov}\{f(X_t),f(X_{t+h})\}
    =\rho^h\operatorname{Var}_p\{f(X)\}
\end{equation}
within a block, giving variance-inflation factor
\begin{equation}
    \tau_L(\rho)
    =1+2\sum_{h=1}^{L-1}\left(1-\frac{h}{L}\right)\rho^h,
    \qquad
    n_{\rm eff}=\frac{n}{\tau_L(\rho)}.
    \label{eq:neff}
\end{equation}
At $\rho=0.35$, $\tau_L=2.070$ and $n_{\rm eff}=48\,302$. Figure~\ref{fig:neff} compares the Markov estimate as $\rho$ increases with the estimate from an i.i.d. Zipf sample of size $n_{\rm eff}(\rho)$. The curves are close through moderate persistence and separate at high $\rho$. Thus $n_{\rm eff}$ reproduces most of the finite-sample shift over the range relevant to the primary design. We do not treat it as a bias correction: empirical re-ranking, truncation and nonlinear estimation prevent an exact reduction to an i.i.d. sample of size $n_{\rm eff}$.

This separation between a fixed frequency marginal and sequence-dependent finite-sample behavior is consistent with a broader point made recently for the Heaps--Zipf relation: temporal correlations can alter type--token growth even when the rank--frequency distribution is unchanged \cite{zimmerlin2026}. Here the affected observable is different---a fitted rank exponent rather than a type--token curve---but the comparison isolates the same loss of information incurred when temporal order is discarded.

Dependence also predicts the smaller realized Markov vocabulary. The exact expectation derived in Appendix~\ref{app:robustness} is 4598.88 under the primary parameters, compared with a held-out mean of 4601.0; the corresponding i.i.d. values are 4854.88 and 4857.8. The vocabulary difference is therefore a finite-sample consequence of the dependence structure in this construction, not simply an unmatched simulation parameter.

\subsection{Conditioning reveals aggregation in the latent mixture}

\begin{figure}[t]
\centering
\includegraphics[width=0.74\textwidth]{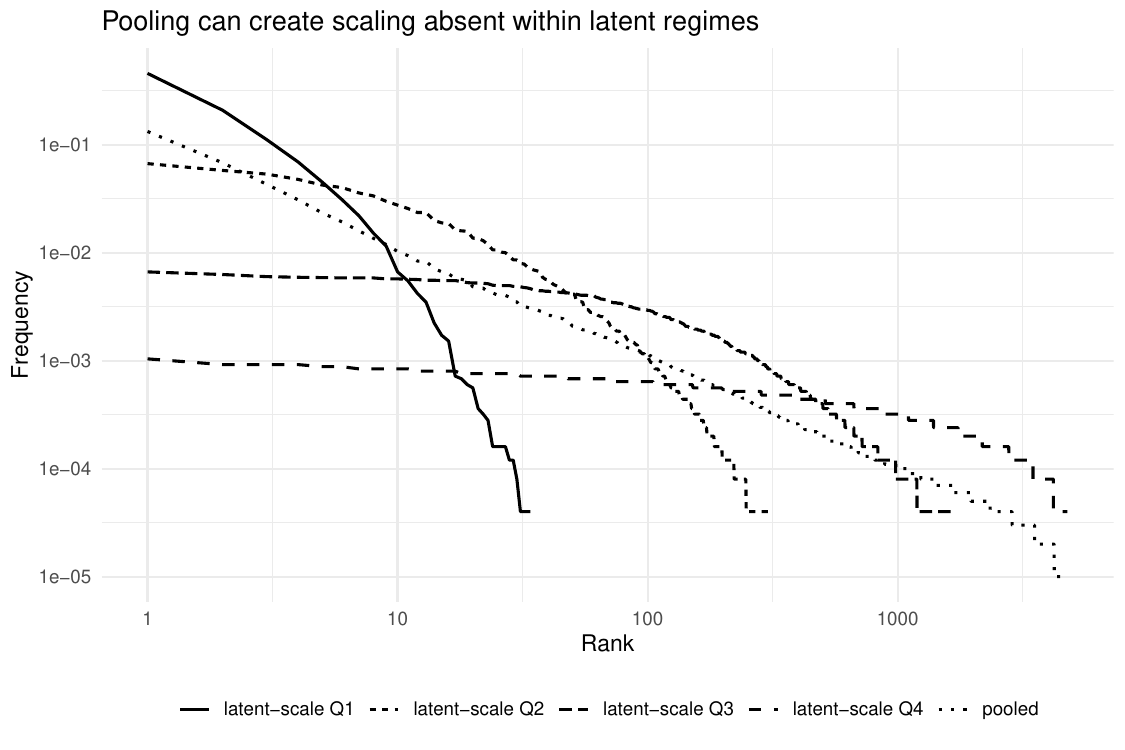}
\caption{Rank--frequency curves for quartiles of the latent scale and for the pooled sample. Pooling produces the approximate scaling used in the marginal comparison even though the conditional rank distributions are truncated exponentials.}
\label{fig:latent}
\end{figure}

The latent mixture produces the marginal pattern by a different route. Conditional on its sequence-specific scale, observations are independent and follow a truncated exponential law. Figure~\ref{fig:latent} groups sequences by scale quartile. The pooled curve is close to linear over the primary rank domain, while the stratum-specific curves differ substantially in shape and support.

For this construction, conditioning exposes the source of the apparent law. In empirical data the same exercise would be informative only when the strata correspond to measured or substantively plausible sources of heterogeneity. A Zipf-like pattern that persists within such strata would weaken, but not eliminate, an aggregation explanation.

\subsection{Order information and transition direction}
\label{sec:sequence}

Sequence statistics are computed only for transitions that remain within the declared sequence boundaries. The $K=50$ most frequent empirical types are retained and all remaining types are collapsed into an \textsc{other} category. We define
\begin{equation}
    I_{\rm NMI}(X;Y)
    =\frac{I(X;Y)}{\sqrt{H(X)H(Y)}}
\end{equation}
and
\begin{equation}
    I_{\rm excess}
    =I_{\rm NMI}(X_t;X_{t+1})
    -\E_{\rm shuffle}\!\left[I_{\rm NMI}(X_t;X_{t+1})\right],
    \label{eq:iexcess}
\end{equation}
where the shuffle mean is estimated from 20 independent permutations within each sequence. Subtracting this baseline preserves sequence lengths and within-sequence marginal frequencies while removing order.

For direction, let $R_t$ denote the global empirical frequency rank of token $t$ and set
\begin{equation}
    D=\Prob(R_{t+1}<R_t\mid R_{t+1}\neq R_t).
    \label{eq:direction}
\end{equation}
Because rank 1 denotes the most frequent type, $R_{t+1}<R_t$ is a move toward a more frequent empirical type. $D$ is a statistic of the observed sequence, not a direct measurement of the transition rule. The two sequential controls nevertheless make analytic predictions about it. For the Markov chain, off-diagonal stationary flows satisfy $p_iP_{ij}=p_jP_{ji}$, so upward and downward rank-changing transitions are equally likely for any fixed ordering and the population value is $D=1/2$. In canonical SSR, state order coincides with stationary frequency order and every within-cycle transition moves to a lower state index. With theoretical ranks this gives $D=1$; empirical re-ranking and the treatment of sequence boundaries reduce the observed value.

\begin{figure}[t]
\centering
\includegraphics[width=0.94\textwidth]{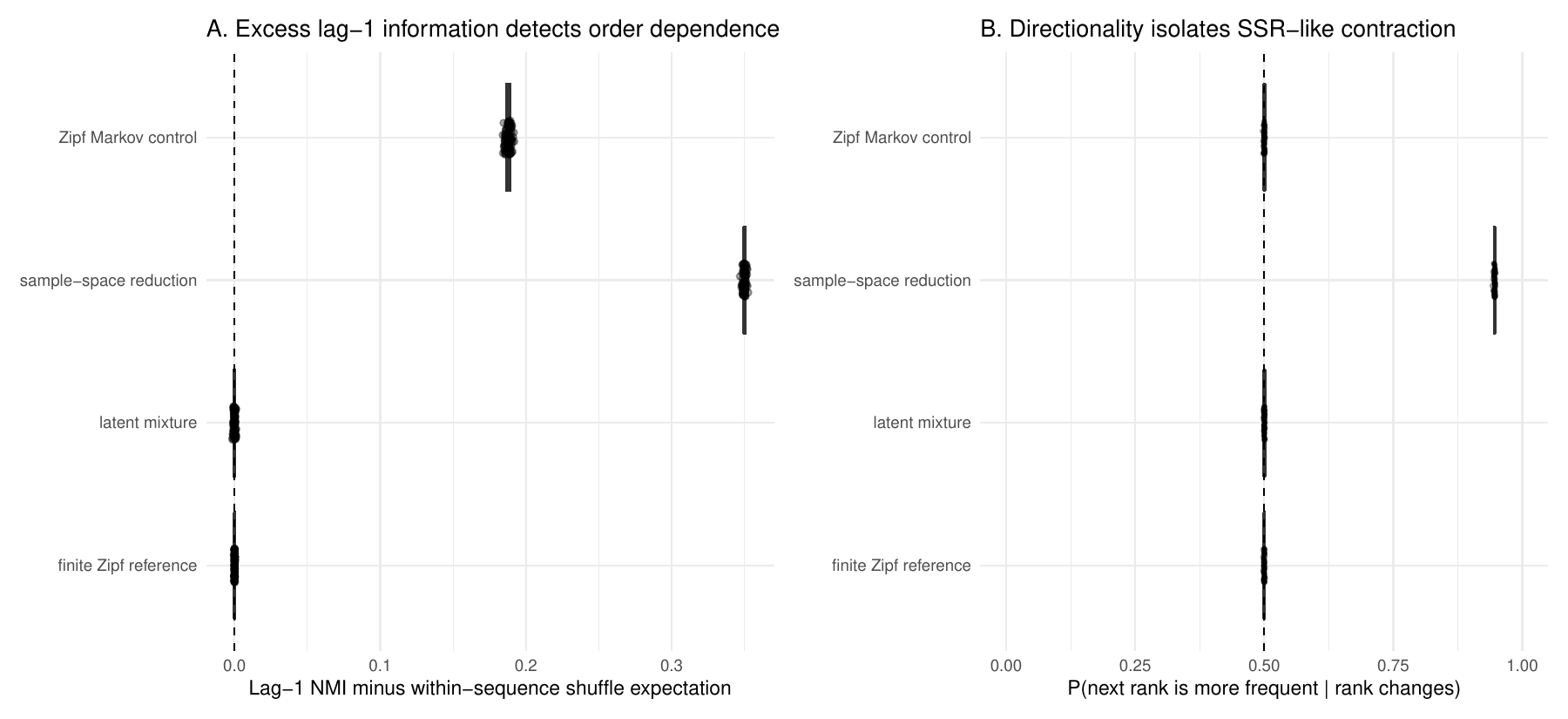}
\caption{Sequence statistics after the marginal comparison. Panel A reports excess lag-1 normalized mutual information relative to within-sequence shuffles. Panel B reports the fraction of rank-changing transitions that move toward a more frequent empirical type. Markov persistence and SSR both contain order information, while direction distinguishes the symmetric rank flow of the reversible Markov control from canonical SSR contraction.}
\label{fig:sequence}
\end{figure}

The simulations reproduce these contrasts (Figure~\ref{fig:sequence}). Mean $I_{\rm excess}$ is approximately zero for the i.i.d. reference and latent mixture, 0.188 for Markov and 0.350 for SSR. Thus excess lag-1 information detects order dependence relative to the shuffle null but does not distinguish its form. Direction provides the second contrast: $D=0.500$ for Markov and 0.946 for SSR. The large SSR value is a positive control for the contraction built into the model, not evidence discovered after examining the simulations.

The AUC effects in Appendix Figure~\ref{fig:bootstrap} reach the boundary for excess NMI when an order-dependent construction is compared with an exchangeable one, and for $D$ when SSR is compared with the other models. These near-trivial positive controls are intentional: they show what is gained by measuring a feature that the candidate processes are known to differ on. They do not establish either statistic as a unique empirical signature.

The sequence results are stable to two prespecified perturbations. As the retained alphabet increases from $K=25$ to 100, mean $I_{\rm excess}$ remains near zero for the i.i.d. reference and latent mixture, rises from 0.176 to 0.198 for Markov, and falls from 0.378 to 0.319 for SSR. Replacing natural SSR cycles with fixed 250-token blocks changes $D$ from 0.946 to 0.842 and $I_{\rm excess}$ from 0.350 to 0.358. The magnitude of directional asymmetry is therefore boundary-dependent, while the qualitative distinction remains over the two boundary definitions considered here.

\subsection{Fit-window sensitivity and observation-induced mimicry}
\label{sec:windows}

The fitted exponent also depends on the rank domain. Holding all generator parameters fixed, we vary
\[
    r_{\max}\in\{300,500,1000,2000,3000\}
\]
and
\[
    r_{\min}\in\{1,5,10,20,50,100\},
\]
with the count rule $N_{(r)}\geq5$ unchanged. Appendix Figures~\ref{fig:rmax} and \ref{fig:rmin} report the results.

Two auxiliary constructions make the observation-window issue explicit. In the random-segmentation control, independent letters are drawn from a fixed unequal alphabet and concatenated into positive-length units whose lengths follow a delimiter-induced geometric law. Simon reinforcement introduces a new type with fixed innovation probability and otherwise resamples a previous token, giving reuse probability proportional to current abundance. Their parameters are calibrated only for this fit-window exercise and are recorded in the reproducibility archive.

Random segmentation is especially sensitive to the upper cutoff. Its median OLS exponent is near 1 at the primary $r_{\max}=500$, about 1.2 at $r_{\max}=300$, and about 1.19 once larger requested windows become count-limited. Simon reinforcement remains close to one over the same range. This does not define a categorical boundary between ``compatible'' and ``mimicking''; it shows that the near-one regime for random segmentation is tied to the observation window.

The lower-cutoff sweep is restricted to the four primary constructions because its purpose is to test the robustness of the main marginal comparison, rather than the auxiliary observation-window examples. It mainly affects the latent mixture. Its mean RD-MLE falls from 1.048 at $r_{\min}=1$ to 1.004 at $r_{\min}=10$ and 0.998 at $r_{\min}=20$, reflecting curvature in the head of the pooled distribution. The i.i.d. and SSR estimates remain much closer over these cutoffs. We retain $r_{\min}=10$ because it excludes the strongest head deviation while preserving the primary marginal comparison; the sensitivity figure makes that choice explicit rather than treating it as an estimated cutoff.

These experiments separate three questions that a fitted exponent alone conflates: whether a pattern is created by pooling, whether observations carry serial dependence, and whether that dependence is directional. Which question can be answered depends on the observation design. A mechanistic interpretation therefore requires a model for both the data-generating process and the operations that turn it into the analyzed sample.

\section{Implications for empirical sequence data}
\label{sec:animal}

The benchmark suggests a general strategy for empirical sequence data: after establishing a marginal scaling pattern, retain or collect observables that probe temporal organization, heterogeneity, or dynamics. This is closely related to recent work showing that temporal correlations can break the usual Heaps--Zipf correspondence in human discovery sequences \cite{zimmerlin2026}.

Learned animal communication provides a concrete biological case. Humpback whale song is culturally transmitted, and statistical segmentation identified recurrent parts with a strongly skewed frequency distribution \cite{arnonWhale2025}. Bengalese finch song is socially learned, and recent work reports coherent subsequences with a Zipfian rank--frequency pattern \cite{kirbyBirdsong2026}. We do not reanalyze either dataset. The question is which additional observations would distinguish explanations once the marginal scaling pattern has been established.

Table~\ref{tab:animalpredictions} lists several such contrasts. None is unique to a single mechanism, and several can hold at the same time.

\begin{table}[t]
\centering
\footnotesize
\caption{Candidate contrasts for learned animal communication, used here as one example of empirical sequence data. Interpretations are comparative and non-exclusive; cultural transmission can operate through reinforcement, coding or transition constraints.}
\label{tab:animalpredictions}
\begin{tabularx}{\textwidth}{>{\raggedright\arraybackslash}p{4.3cm} X}
\toprule
\textbf{Observed pattern} & \textbf{Interpretation supported} \\
\midrule
Pooled Zipf-like scaling weakens after conditioning on individual, tutor, session, sequence length or another measured state & Latent heterogeneity contributes to the marginal pattern; aggregation is substantively important. \\
\addlinespace
Rank--frequency scaling survives conditioning and $I_{\rm excess}$ is positive relative to a within-sequence shuffle null & The sequence contains order information not represented by the marginal distribution. \\
\addlinespace
Positive excess transition information with approximately symmetric rank direction ($D\approx1/2$) & Temporal dependence without evidence, from this statistic, for directional sample-space contraction. \\
\addlinespace
Positive excess transition information together with strong directional or nested contraction of accessible successors & Support for SSR-like transition restrictions, conditional on the segmentation and sequence-boundary definitions. \\
\addlinespace
Current tutor or learner frequency predicts subsequent frequency increments after accounting for exposure and age & A reinforcement or cumulative-advantage component is compatible with the trajectory. \\
\addlinespace
Statistics change systematically across tutor--learner links, development or transmission generations & Transmission or learning contributes to the trajectory rather than merely describing a pooled endpoint. \\
\addlinespace
Frequency covaries with duration, energetic cost, predictability or information in the direction predicted by a coding model & Evidence relevant to an efficient-coding or communicative-optimization account. \\
\bottomrule
\end{tabularx}
\end{table}

Segmentation must remain part of the observation model. If unit boundaries are estimated from the same sequence structure later used for diagnosis, preprocessing can induce, suppress or condition the dependence measured by $I_{\rm excess}$ and $D$. Surrogates and conditioning analyses should therefore repeat the segmentation step whenever the segmentation algorithm is itself data-dependent, rather than treating inferred units as fixed observations. Sequence statistics also need a hierarchical interpretation: positive $I_{\rm excess}$ shows order dependence relative to the chosen shuffle null, whereas claims about SSR require additional evidence on direction, nested successor sets or longer-memory restrictions. Tutor--learner and developmental data address a different level of explanation because transmission accounts concern trajectories, not only endpoints.

Real communication data add measurement error, incompletely observed context, uncertain boundaries and potentially several concurrent mechanisms. Table~\ref{tab:animalpredictions} should therefore be read as a set of relative compatibility and falsification contrasts, not as a point-identification scheme. The simulations provide controlled reference cases for designing such tests; they are not models calibrated to whale or birdsong production.

Cultural transmission can also sit upstream of the lower-level mechanisms in Table~\ref{tab:animalpredictions}. Learning may alter reuse probabilities, coding costs or the set of accessible successors, and those changes can in turn alter the rank--frequency distribution. Longitudinal or tutor--learner data are needed to distinguish these links from a statistical regularity observed only after pooling.

\section{Discussion: from scaling laws to mechanisms}

The controlled comparison separates three objects that are often collapsed in empirical scaling analyses: a population marginal, the sampling distribution of a fitted exponent, and temporal organization. The i.i.d. reference, persistent Markov chain, and canonical SSR have the same stationary rank distribution, yet their joint dynamics differ. The Markov example adds a finite-sample effect: even with an identical population marginal, dependence changes the distribution of the fitted exponent. The effective-sample-size calculation accounts for most of this shift at moderate persistence, while the residual discrepancy makes clear that empirical ranking and truncation cannot be reduced exactly to an i.i.d. sample-size correction.

The latent mixture illustrates a separate failure of one-to-one interpretation. Its conditional rank distributions are truncated exponentials, but pooling over scales produces a Zipf-like marginal. Conditioning can therefore change the evidential content of the scaling law rather than merely check its robustness. Sequence perturbations serve a parallel purpose for temporal structure. In the simulations, excess lag-1 information detects order dependence and transition direction separates two forms of that dependence because the candidate processes make different predictions for those quantities.

The same design principle extends to other mechanism classes. Growth increments probe reinforcement or proportional growth; frequency--cost relations probe coding accounts; finite-size and response statistics are needed for claims about criticality; and longitudinal changes probe transmission or learning. Recent results on dependency structures, temporal component statistics, and Heaps--Zipf decoupling make the same point from complementary directions \cite{mazzolini2018dependency,mazzolini2018ssr,zimmerlin2026}: temporal or relational observables can carry information that is absent from a one-point frequency law. Such evidence can rule out classes of explanations or establish that a structural feature is required even when no single mechanism is point identified.

\subsection*{Scope and limitations}

The simulation study covers four primary constructions from a broader taxonomy. The latent mixture, Markov chain, and canonical SSR are minimal reference models, and the exact marginal equivalence proved here applies to these specific finite-state constructions. Multiple mechanisms can operate simultaneously in empirical systems. The sequence statistics also depend on how the alphabet and sequence boundaries are defined; the sensitivity analyses establish robustness only over the perturbations examined. Learned communication is included as an empirical case study, not as a calibrated application: cultural transmission is not implemented as a generator in the benchmark, and no biological recordings are reanalyzed.

A fitted scaling law constrains candidate mechanisms but cannot select among mechanisms that reproduce the same marginal information. Discrimination requires additional observables tied to predictions on which those candidates differ.

\appendix
\section{Calibration, analytical checks and robustness}
\label{app:robustness}

\subsection{Calibration and reproducibility}

Only tunable constructions are calibrated. For the latent mixture, a broad grid is first screened on 10 calibration seeds. The leading candidates and local refinements are then evaluated on the prespecified 50-seed panel 1001--1050. Candidates are ranked by $|\overline{\widehat\alpha}_{\rm OLS}-1|$, with RMSE relative to one and the Monte Carlo standard deviation used as secondary criteria. The selected scale interval is $[0.5,8500]$. Random segmentation and Simon reinforcement are calibrated separately for the fit-window exercise. No conditioning or sequence statistic enters any calibration criterion, and held-out seeds beginning at 2001 are not used for parameter selection. The R/Quarto workflow records the full candidate grids and seeds.

On the calibration samples, the selected latent mixture has mean OLS exponent 1.001 with standard deviation 0.033. Random segmentation and Simon reinforcement have calibration means 1.001 and 0.998, respectively. These values are not used as held-out evidence.

OpenAI ChatGPT (GPT-5.6 Sol, August 2026) was used during development and debugging of parts of the R/Quarto workflow. All AI-assisted code affecting reported results was inspected and executed by the author, and the analytical checks and numerical outputs were independently verified. Additional use in manuscript preparation is disclosed in the Acknowledgments.

\subsection{Markov persistence and expected vocabulary}

Table~\ref{tab:rho} gives the persistence analysis underlying Figure~\ref{fig:neff}. Each row is based on 100 independent replications from a seed block separate from the primary held-out benchmark, so the entry at $\rho=0.35$ need not equal the corresponding mean in Table~\ref{tab:simresults}. The effective-sample-size comparison follows the Markov exponent shift closely through moderate $\rho$ and becomes less accurate at high persistence.

The realized vocabulary can also be calculated under the Markov construction. For type $j$, the probability of no occurrence in a block of length $L$ is
\begin{equation}
    q_j=(1-p_j)\{1-(1-\rho)p_j\}^{L-1}.
\end{equation}
With $B=n/L$ independent blocks,
\begin{equation}
    \E[V_n^{\rm Markov}]
    =\sum_{j=1}^{V}\left(1-q_j^B\right),
    \qquad
    \E[V_n^{\rm iid}]
    =\sum_{j=1}^{V}\left[1-(1-p_j)^n\right].
    \label{eq:expected-vocab}
\end{equation}
Under the primary parameters the expectations are 4598.88 and 4854.88, respectively. They are close to the held-out means 4601.0 and 4857.8.

\begin{table}[H]
\centering
\footnotesize
\caption{Sensitivity to Markov persistence. Each row averages 100 independent replications from a seed block distinct from the primary benchmark. The i.i.d. comparison uses $n=n_{\rm eff}$ from equation~\eqref{eq:neff}; $\E[V_n]$ is the exact expectation from equation~\eqref{eq:expected-vocab}.}
\label{tab:rho}
\begin{tabular}{rrrrrrr}
\toprule
$\rho$ & $n_{\rm eff}$ & Markov OLS & i.i.d. OLS & Markov RD-MLE & i.i.d. RD-MLE & simulated / expected $V_n$ \\
\midrule
0.00 & 100000 & 0.9928 & 0.9930 & 0.9956 & 0.9953 & 4855.2 / 4854.9 \\
0.10 & 81884 & 0.9920 & 0.9913 & 0.9941 & 0.9950 & 4804.5 / 4807.3 \\
0.20 & 66778 & 0.9898 & 0.9898 & 0.9928 & 0.9928 & 4742.5 / 4742.9 \\
0.35 & 48302 & 0.9855 & 0.9861 & 0.9909 & 0.9912 & 4599.8 / 4598.9 \\
0.50 & 33512 & 0.9788 & 0.9797 & 0.9871 & 0.9875 & 4360.8 / 4362.1 \\
0.65 & 21405 & 0.9643 & 0.9706 & 0.9769 & 0.9791 & 3961.6 / 3956.5 \\
0.80 & 11312 & 0.9311 & 0.9517 & 0.9548 & 0.9687 & 3208.2 / 3208.7 \\
\bottomrule
\end{tabular}
\end{table}

\subsection{Pairwise bootstrap effects}

\begin{figure}[H]
\centering
\includegraphics[width=0.92\textwidth]{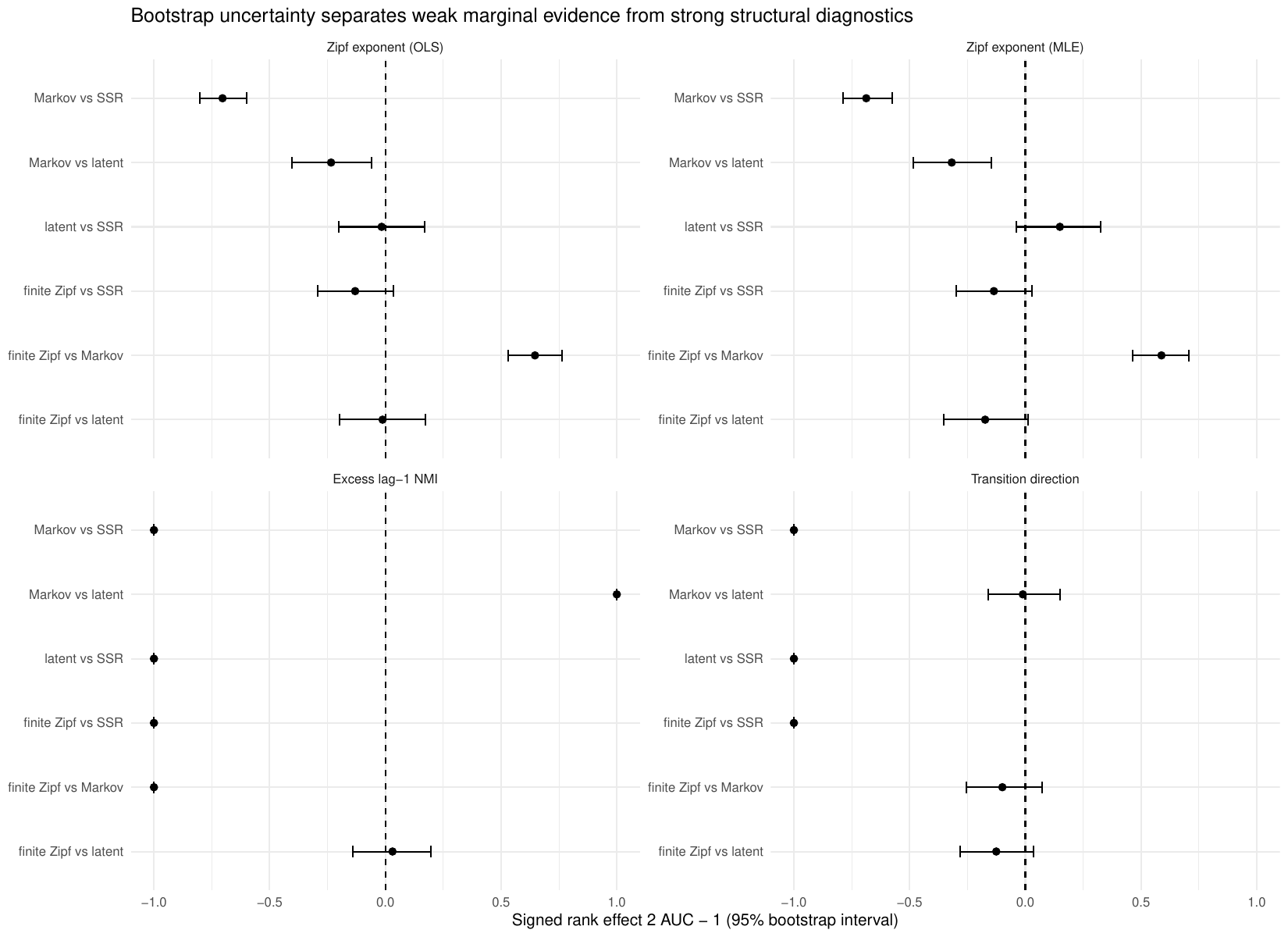}
\caption{Signed pairwise rank effects $2\,\mathrm{AUC}-1$ with 95\% bootstrap intervals. Exponent effects are small for finite Zipf versus latent mixture and for comparisons involving SSR, while the Markov finite-sample shift is visible despite its identical stationary marginal. Excess NMI separates order-dependent from exchangeable constructions in these simulations, and transition direction separates SSR from the other models. The sign records ordering and the magnitude records univariate discrimination.}
\label{fig:bootstrap}
\end{figure}

\subsection{Alphabet and sequence-boundary sensitivity}

\begin{figure}[H]
\centering
\begin{subfigure}[t]{0.48\textwidth}
\centering
\includegraphics[width=\linewidth]{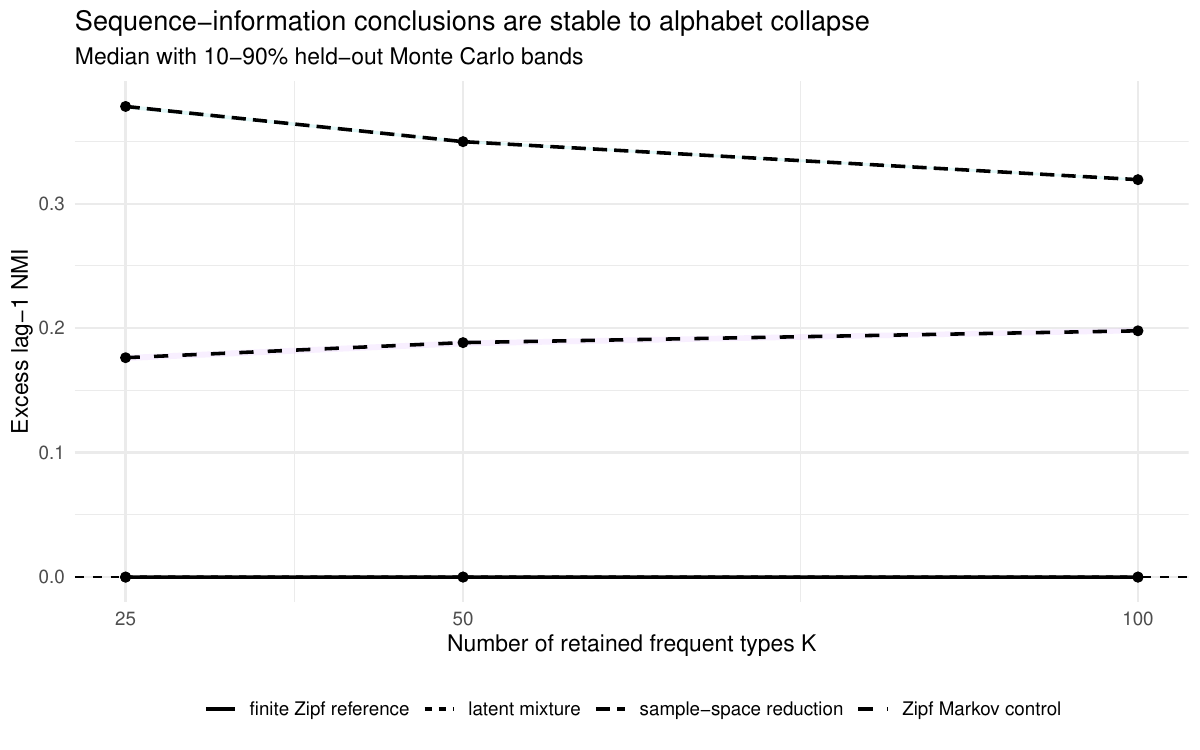}
\caption{Alphabet-collapse sensitivity.}
\end{subfigure}\hfill
\begin{subfigure}[t]{0.48\textwidth}
\centering
\includegraphics[width=\linewidth]{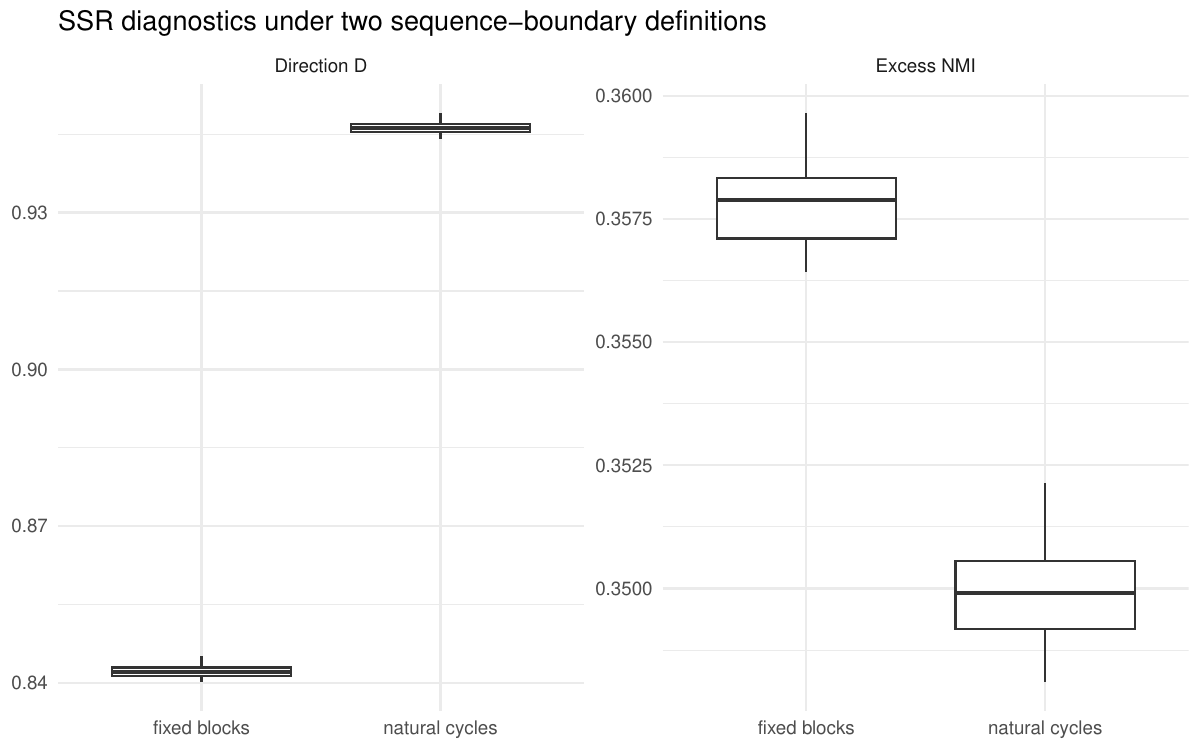}
\caption{SSR sequence-boundary sensitivity.}
\end{subfigure}
\caption{Sensitivity of the sequence statistics. Panel A varies the number $K$ of frequent types retained before collapsing the tail into \textsc{other}. Panel B compares natural SSR cycles with fixed 250-token blocks. The ordering is unchanged over the values examined, although the magnitude of $D$ depends on the boundary definition.}
\label{fig:sequence-robustness}
\end{figure}

\subsection{Rank-window sensitivity}

\begin{figure}[H]
\centering
\includegraphics[width=0.84\textwidth]{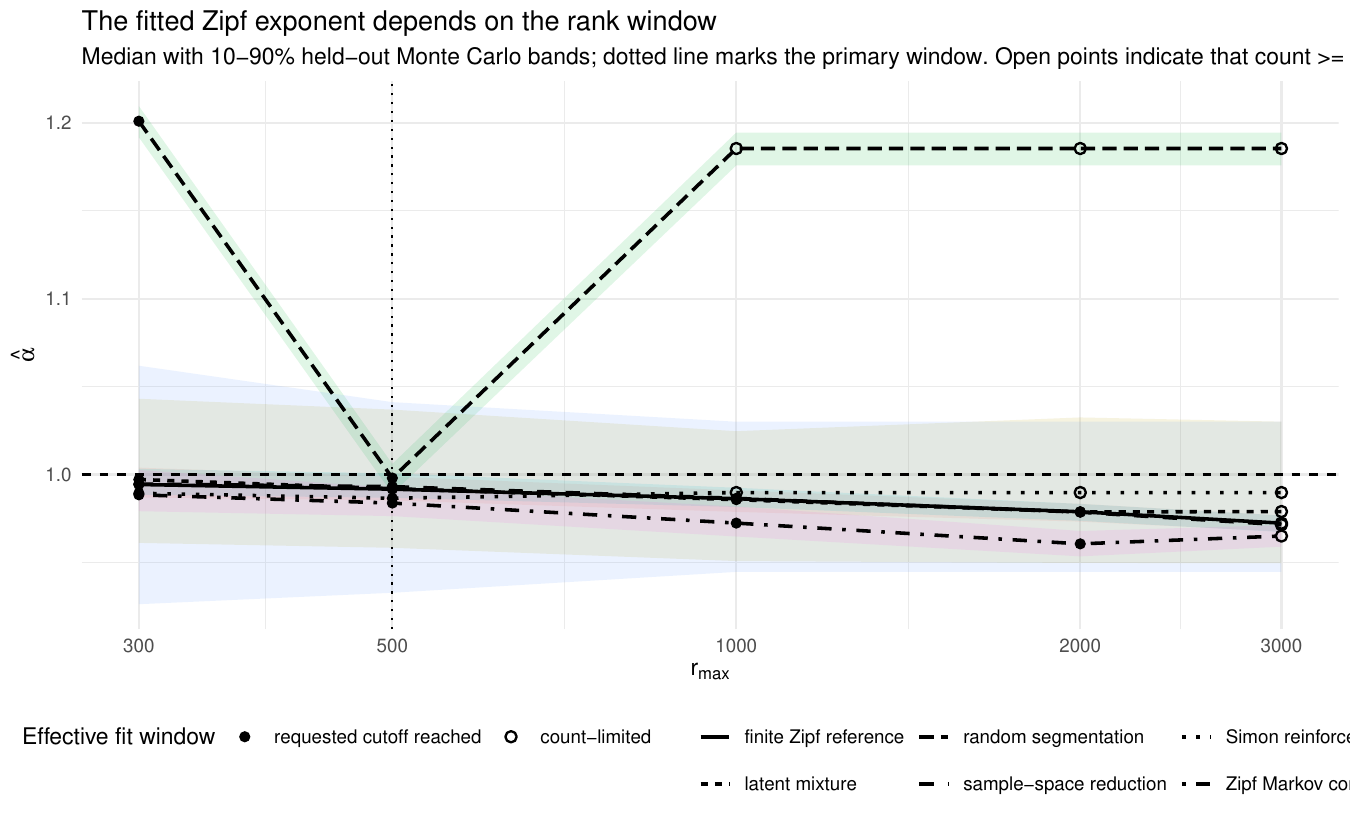}
\caption{Sensitivity of the fitted OLS exponent to the requested upper rank. Curves show medians and 10--90\% held-out Monte Carlo bands; the vertical dotted line marks the primary $r_{\max}=500$ window. Open points indicate that the count threshold, rather than the requested cutoff, determines the effective upper rank. Parameters are fixed throughout.}
\label{fig:rmax}
\end{figure}

\begin{figure}[H]
\centering
\includegraphics[width=0.84\textwidth]{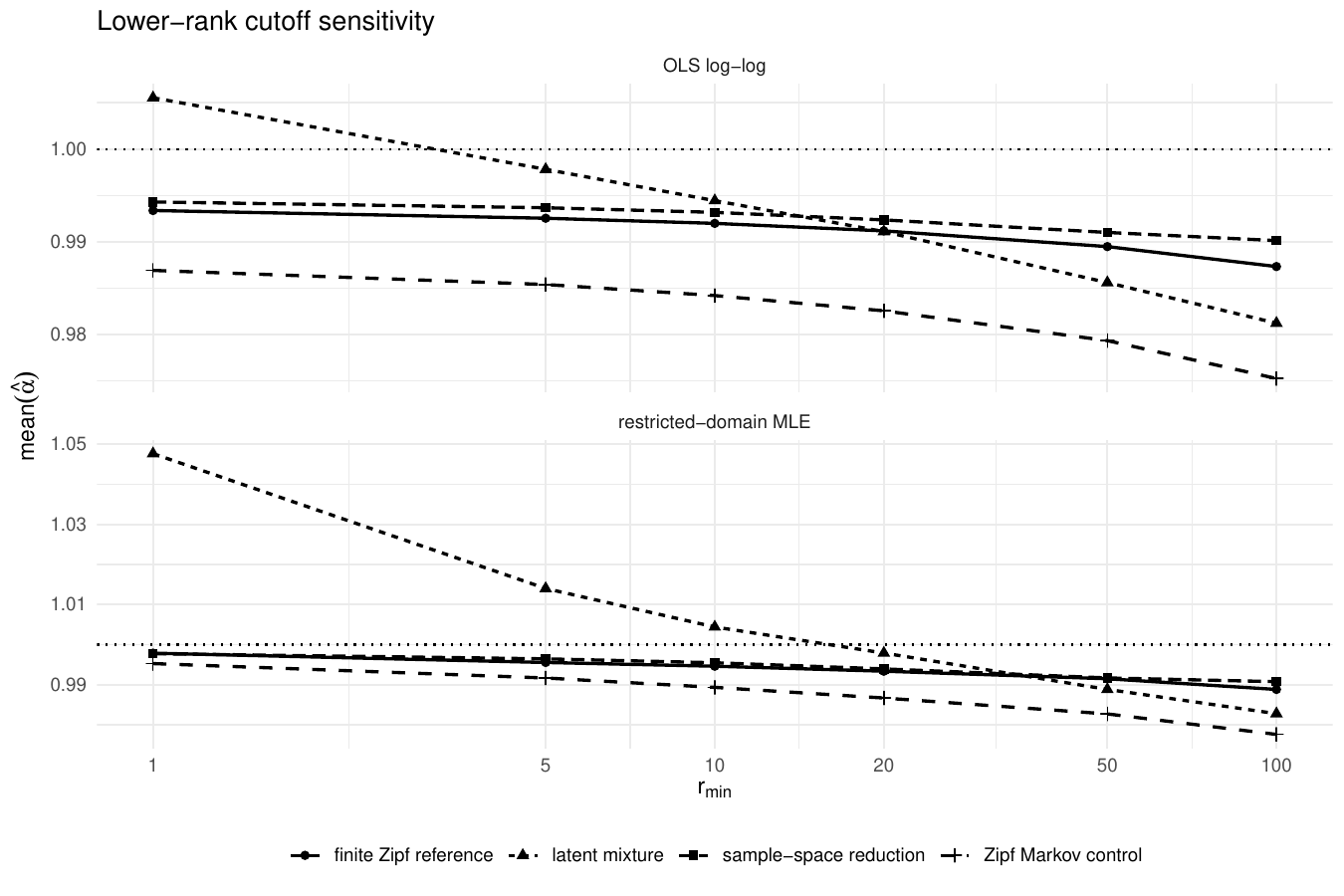}
\caption{Sensitivity to the lower fitted rank for the four primary constructions. The primary value $r_{\min}=10$ avoids the largest head deviations of the latent mixture while leaving the i.i.d. and SSR comparison nearly unchanged over moderate cutoffs. Both OLS and RD-MLE are shown.}
\label{fig:rmin}
\end{figure}

\printbibliography

\end{document}